\documentclass{optica-article}

\journal{opticajournal} 

\articletype{Research Article}

\usepackage{lineno}

\usepackage{amsmath,bm}
\usepackage{physics}
\usepackage{wrapfig}
\usepackage{subcaption}

\renewcommand{\vec}[1]{\mathbf{#1}}

\begin{document}

\title{Dynamic Mueller matrix polarimetry using generalized measurements}

\author{Amy McWilliam,\authormark{1,*} Mustafa A. Al Khafaji,\authormark{1,2} Sphinx J. Svensson,\authormark{1} Sebasti\~ao P\'adua,\authormark{3}  and Sonja Franke-Arnold\authormark{1}}

\address{\authormark{1}School of Physics and Astronomy, University of Glasgow, G12 8QQ, Glasgow, Scotland, UK\\
\authormark{2}Fraunhofer CAP, G1 1RD, Glasgow, Scotland, UK\\
\authormark{3} Departamento de Física, Universidade Federal de Minas Gerais, 31270-901 Belo Horizonte, Minas Gerais, Brazil}

\email{\authormark{*}a.mcwilliam.1@research.gla.ac.uk}

\begin{abstract*}
Mueller matrices provide a complete description of a medium's response to excitation by polarized light, and their characterization is important across a broad range of applications from ellipsometry in material science to polarimetry in biochemistry, medicine and astronomy. Here we introduce single-shot Mueller matrix polarimetry based on generalized measurements performed with a Poincaré beam. We determine the Mueller matrix of a homogeneous medium with unknown optical activity by detecting its optical response to a Poincaré beam, which across its profile contains all polarization states, and analyze the resulting polarization pattern in terms of four generalized measurements, which are implemented as a path-displaced Sagnac interferometer.
We illustrate the working of our Mueller matrix polarimetry on the example of tilted and rotated wave plates and find excellent agreement with predictions as well as alternative Stokes measurements. After initial calibration, the alignment of the device stays stable for up to 8 hours, promising suitability for the dynamic characterization of Mueller matrices  that change in time.

\end{abstract*}

\section{Introduction}

The polarization of light contains information regarding a light beam's source as well as any interaction with materials that it has encountered. Acquiring polarization information can be achieved with non-invasive techniques, and it is no surprise that polarization profiling is playing a crucial role in many metrological applications. Examples include the characterization and stress analysis of materials \cite{Stifter2003, HEARN1997166, arteagafilms}, ellipsometry \cite{fujiwara2007Ellipsometry, Johnson2000}, astronomy \cite{Hough2006,Lites2008,Costa2001}, pharmaceutical ingredient analysis \cite{Syroeshkin2018}, monitoring of soil conditions and crop growth \cite{Ignatenko2022}, chiral symmetry \cite{arteagachiral}, biomedical studies and clinical  applications \cite{MartinJ2021, Ghosh2011, Ramella-Roman_2020, He2019, AHMAD2020, Matthew2000,He2021, arteagatissues, Tuchin2016}, biological microscopy \cite{Alali2015,Gottlieb:21} and quantum optics and quantum information \cite{goldberg, klyshko, luis, goldberg22, kwiat99,bouwmeester}.

Mueller matrix polarimetry is commonly used to determine the optical activity of an unknown sample (such as its linear/circular birefringence or dichroism) by measuring the polarization changes it induces on a probe beam after transmission or reflection. As any fully or partially polarized state of light can be expressed in terms of a $4\times1$ Stokes vector $\vec{S}$, the optical behavior of an optical element or sample is fully described by a $4\times4$ real Mueller matrix. To determine the elements of the Mueller matrix requires at least 16 measurements taken with linearly independent combinations of settings of the polarization state generator before the sample and the analyzer after the sample.

Polarimetry techniques based on spatial splitting divide the beam and analyze each part using different polarization optics. Alternatively, polarimetry can be achieved by temporal modulation which requires sequential measurements, such as the rotating wave plate approach, which is the technique most commonly found in commercial polarimeters. Spatial splitting techniques tend to be bulkier and deliver inferior signal to noise ratios, making them less suitable for applications requiring weak probe light. Temporal modulation techniques, instead, pose difficulties if the optical activity of the sample varies with time. Dynamic determination of Muller matrices has attracted attention as a tool for studies of rapid phenomena sensitive to polarization \cite{Dahl2001,Aas2010, Li2008,Gramatikov2023,Twietmeyer08,Philpott21}. Noteworthy recent developments include ellipsometers based on photoelastic modulators \cite{Petkovsek2010} which due to their quick switching time allow rapid sequences of individual Stokes measurement, leading to time resolutions in the range of microseconds \cite{Zhang2020}. 

Alternatively, additional degrees of freedom may be utilized to investigate polarization responses, including spectral encoding \cite{Dubreuil07}, where Stokes measurements are simultaneously carried out at different frequency bands, with time resolutions feasible at 10s of nanoseconds \cite{Feng2022}, and spatial encoding, relying on complex vector light, correlated in their spatial and polarization degree of freedom. Polarization information can then be deduced from spatial intensity measurements \cite{Töppel_2014,Hawley2019,Zhang2022}. For a homogeneous sample, it becomes possible to obtain a Mueller matrix using only one probe beam, as long as it contains at least 4 linearly independent polarization states that together span the complex polarization space. An extreme example of such beams are Poincar\'e beams, which contain every possible polarization state across their transverse profile \cite{Beckley}. 
Two succinct examples of Poincar\'e based polarimetry were recently demonstrated: Ref.~\cite{Suarez-Bermejo2019} analyzed 4 regions of the beam, corresponding to 4 linearly independent polarization states with a commercial polarimeter, whereas \cite{Suarez-Bermejo2022} acquired the polarization information from the entire beam with a CCD camera. While these attempts managed to reduce the number of probe beam settings, they still required measurements in multiple analyzer settings.

In this paper we demonstrate one-shot polarimetry by analyzing the response of a homogeneous sample to a Poincar\'e probe beam in terms of simultaneous positive operator valued measurements (POVMs). 
Our method builds on recent work by some of the authors, which demonstrated the single-shot characterization of vector beams by generalized measurements which were performed simultaneously with a Mach-Zehnder type interferometric setup \cite{Mustafa2022}. Here, we propose and demonstrate a significantly improved design for POVM measurements based on a Sagnac interferometer, which can operate stably for timescales in the order of hours without need for re-calibration. 
In combination with a full Poincar\'e beam as the probe, this allows us to obtain the Mueller matrix of an unknown sample in a single shot, enabling dynamic optical activity measurements with potential applications in the investigation of fast physical, chemical or biological processes as well as for stress analysis. The time resolution of the Muller matrix evolution is limited only by the camera temporal resolution. 

We describe the theory of using generalized measurements to identify Mueller matrices in section \ref{sec Theory}, the experimental design of our Sagnac interferometer and its characterization regarding fidelity and stability in section \ref{sec Experiment} and measurements of example Mueller matrices in section \ref{sec Mueller Results}, before offering our conclusions and outlook.

\section{Theory \label{sec Theory}}

\subsection{Generalized measurements \label{sec Generalized measurements}}

Polarization has two degrees of freedom ($d$), and hence requires a minimum of $d^2=4$ measurements for full state reconstruction \cite{James2001,Rehacek,Ling2006}, which are associated with the 4 Stokes parameters. 
Experimentally, it is generally more convenient to measure the Stokes parameters of a polarization state using  six measurements, corresponding to projections into the horizontal, vertical, diagonal, anti-diagonal, right- and left-handed circular polarization bases $(\ket{h},\ \ket{v},\ \ket{d},\ \ket{a},\ \ket{r}$ and $\ket{l})$. Note that our considerations are equally valid in the classical and quantum regime, but for convenience we use a quantum notation throughout.
These projections form an over-complete set of measurements composed from three pairs of mutually unbiased bases.

Whether it is for the determination of the state of a polarization qubit, or for biological applications where light exposure needs to be restricted, tomography using a minimum number of measurements may be desirable. This may be achieved by employing generalized measurements \cite{Nielsen2010QuantumBook,BarnettQuantumInformation,Clarke2001}. 
A commonly used set of operators are those that form a minimum informationally complete positive operator value measure (MIC-POVM) \cite{Renes2004,Paiva-Sanchez2010,Pimenta10,Cardoso2019}. 
The POVM operators $\{\hat{\pi}_i\}$ must be positive ( $\hat{\pi_i}\geq0$ ) and complete ( $\sum_i\hat{\pi_i}=\hat{I}$ where $\hat{I}$ is the identity operator). 
The POVM operators can be written as projection operators $\hat{\pi_i}=\frac{1}{d}\ket{\phi_i}\bra{\phi_i}$, where $|\braket{\phi_i}{\phi_j}|^2=(d\delta_{ij}+1)/(d+1)$. 

\begin{figure}
    \centering    \includegraphics[width=0.92\textwidth]{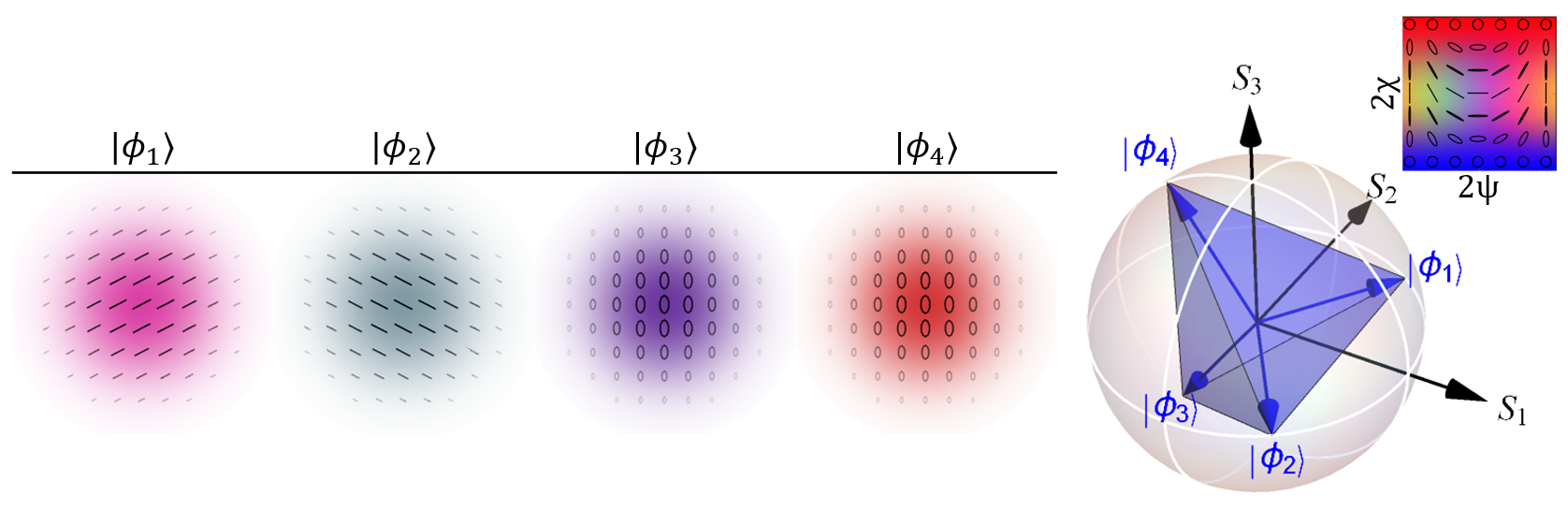}
    \caption{Polarization profiles of the four POVM states described by Eq.~\ref{Eq.POVMStates} and their positions on the Poincar\'e sphere. The states form a tetrahedron on the sphere, as indicated in blue. The color scheme used to represent the polarization is given as an insert on the right, where $\chi$ is ellipticity, $\psi$ orientation angle and the intensity distribution is represented as opacity.}
    \label{fig:POVMStates}
\end{figure}

Utilizing generalized measurements to perform polarization tomography, reduces the number of required measurements from 6 to 4. 
For $d=2$ the states $\ket{\phi_i}$ correspond to the four corners of a tetrahedron lying on the surface of the Poincar\'e sphere with the tetrahedron inserted in the sphere. Their equal spacing and symmetric construction ensures no privilege in reconstructing any polarization state. While the tetrahedron may be oriented arbitrarily on the Poincar\'e sphere, here we chose the geometry indicated in Figure~\ref{fig:POVMStates}, corresponding to POVM states 
\begin{equation}
  \begin{split}
    \ket{\phi_1} = a\ket{h}+b\ket{v}, \ & \ \ket{\phi_2} = a\ket{h}-b\ket{v}, \\
    \ket{\phi_3} = b\ket{h}+ia\ket{v}, \ & \ \ket{\phi_4} = b\ket{h}-ia\ket{v},
  \end{split}
  \label{Eq.POVMStates}
\end{equation}
where 
\begin{equation}
a = \sqrt{\frac{1}{2}+\frac{1}{2\sqrt{3}}}, \quad b = \sqrt{\frac{1}{2}-\frac{1}{2\sqrt{3}}}.
\label{Eq.ab}
\end{equation}
These states have been introduced in \cite{Rehacek,Ling2006} and implemented for spatially varying vector light in \cite{Mustafa2022}. 

For each photon, the probability of measuring an outcome $i$ for a probe state $\ket{\psi}$ is $P_i=\langle \psi |\hat{\pi}_i |\psi \rangle$. For classical light this corresponds to a normalized intensity. Experimentally, we record four intensity profiles $I_i$, from which we obtain the normalized four component vector $\vec{I}^{(\text{N})}= (I_1,I_2,I_3,I_4)/\sum_{i=1}^{i=4}I_i$. The vector $\vec{I}^{(\text{N})}$ is related to the (normalized) Stokes vector by
\begin{equation}
    \vec{I}^{(\text{N})} = \Pi\cdot\vec{S}^{(\text{N})},
    \label{Eq.Projection}
\end{equation}
where $\Pi$ is the $4\times4$ instrumentation matrix \cite{Ling2006}. For an ideal experimental setup this is given by
\begin{equation}
\label{Eq ideal Pi}
    \Pi = \frac{1}{4}
    \begin{pmatrix}
      1 & \sqrt{\frac{1}{3}} & \sqrt{\frac{2}{3}} & 0 \\
      1 & \sqrt{\frac{1}{3}} & -\sqrt{\frac{2}{3}} & 0 \\
      1 & -\sqrt{\frac{1}{3}} & 0 & -\sqrt{\frac{2}{3}} \\
      1 & -\sqrt{\frac{1}{3}} & 0 & \sqrt{\frac{2}{3}}
    \end{pmatrix}.
\end{equation}
The horizontal entries of $\Pi$ correspond to the Stokes vectors of each of the POVM states, $\ket{\phi_i}$. By inverting Eq.~\ref{Eq.Projection}, Stokes vectors can be recovered from four intensity measurements \cite{Ling2006,Mustafa2022}. In this work we are interested in analyzing complex vector light featuring spatially varying polarization profiles, e.g.~ Poincar\'e beams, before and after they have interacted with a sample. Consequently, by realizing generalized measurements with the POVM states results in spatially varying intensity profiles $\vec{I}^{(\text{N})}(x,y)$, corresponding to spatially varying Stokes vectors $\vec{S}^{(\text{N})}(x,y)$.

\subsection{Mueller matrix determination using a full Poincar\'e beam \label{sec Mueller Theory}}

The optical behavior of any optical element or unknown sample can be described by a $4\times4$ Mueller matrix $\vec{M}$, such that an initial polarization state $\vec{S}_{\rm in}$ is transformed into $\vec{S}_{\rm out} = \vec{M}\,\vec{S}_{\rm in}$ after passing through the sample  \cite{Rogovtsov2009}. While this relation is conventionally applied to homogeneously polarized light, it applies equally to spatially varying vector light so that $\vec{S}_{\rm out}(x,y) = \vec{M}\,\vec{S}_{\rm in}(x,y)$.  
We have outlined in section \ref{sec Generalized measurements} how to analyze Stokes vectors simultaneously in the POVM strategy.

One method of obtaining the 16 elements of $\vec{M}$ would be to illuminate the sample with light that contains polarizations corresponding to all four POVM states, and then analyzing it in the POVM outputs. Here we will use Poincar\'e beams as the probe. These beams contain all polarizations, including those of the chosen POVM states (and in fact those of any rotated POVM tetrahedron). In order to maximize the information obtained from our measurement, we use all polarizations of the probe beam, and analyze its components in the POVM interferometer. 

While intensity profiles and corresponding Stokes vectors are defined for continuous spatial coordinates $(x,y)$, their measurement with CCD cameras suggests parametrization by pixel number. 
For $N$ camera pixels used to measure the input and output beams, the measured Stokes vectors can be arranged into $4 \times N$ matrices $\mathcal{S}$, with $\mathcal{S}_{i,n}$ denoting pixel $n \in \{1, \dots N\}$ and parameter $i \in \{1,2,3,4\}$. This allows us to obtain a set of  $4N$ linear equations:
\begin{equation}
    \mathcal{S}_{\rm out} = \vec{M}\,\mathcal{S}_{\rm in},
    \label{Eq.LinearEquations}
\end{equation}
where the matrices $\mathcal{S}_{\rm in}$ and $\mathcal{S}_{\rm out}$ are obtained from intensity measurements by implementing the POVM. 
To solve Eq.~\ref{Eq.LinearEquations} for $\vec{M}$ requires us to find the inverse for $\mathcal{S}_{\rm in}$, which is a non-square matrix. We follow a similar procedure as outlined in \cite{Suarez-Bermejo2022}, and calculate the right Moore-Penrose pseudo-inverse of $S_{\rm in}$ to be calculated via \cite{Ben-Israel},  
\begin{equation}
    (\mathcal{S}_{\rm in})^{\dagger} = \left(\mathcal{S}_{\rm in}\right)^T \left[ \mathcal{S}_{\rm in}\left(\mathcal{S}_{\rm in}\right)^T \right]^{-1}.
    \label{Eq.Moore-Penrose}
\end{equation}
Combining Eqs.~\ref{Eq.LinearEquations} and~\ref{Eq.Moore-Penrose}, the Mueller matrix can be obtained using,
\begin{equation}
    \vec{M} = \mathcal{S}_{\rm out}(\mathcal{S}_{\rm in})^{\dagger}, 
    \label{Eq.SolveMueller}
\end{equation}
benefiting from the full information contained in the Poincar\'e beams which can be analyzed simultaneously in the 4 POVM states. We note that the Moore-Penrose method works only if the beam profile contains 4 linearly independent polarization states.

\section{Experimental design \label{sec Experiment}}

\subsection{Sagnac interferometer for spatial POVM tomography \label{sec setup}}

The experimental setup for Mueller matrix measurements consists of the beam generation stage to produce Poincar\'e beams, an optional interaction with the sample, and a single-shot polarization tomography based on POVM measurements. 

Various methods are available to generate Poincar\'e beams, in our experiment we use an interferometric setup containing a digital mirror display (DMD) to shape two orthogonal polarization components of the input light independently, following the methods introduced in \cite{Selyem2019,Rosales-Guzman2020}. 

\begin{figure}[ht]
    \centering    \includegraphics[width=0.85\linewidth]{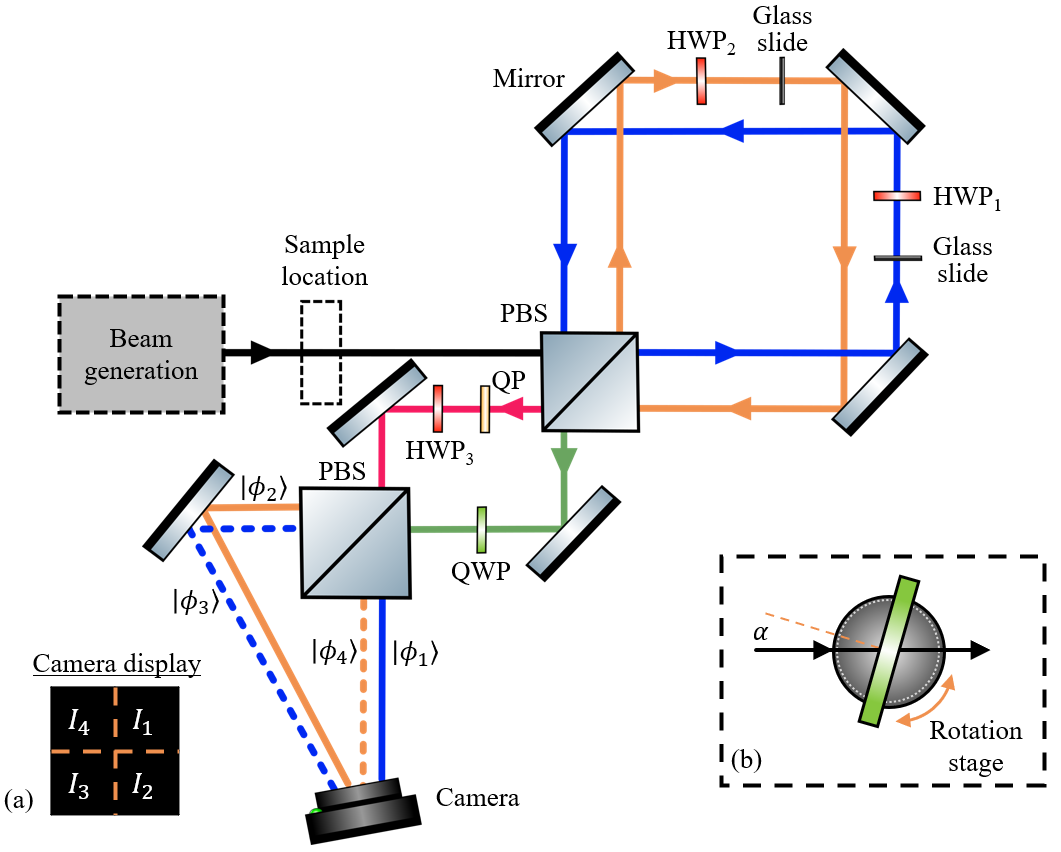}
    \caption{Experimental design for single-shot, spatially dependent polarization tomography and Muller matrix determination using POVM measurements. PBS: polarizing beam splitter, HWP: half waveplate, QWP: quarter waveplate, QP: quartz plate. (a) The arrangement of the four quadrants on the camera display. (b) QWP on a rotation stage used for the tilted QWP measurements in section \ref{TiltedQWPSection}.}
    \label{fig:SetUp}
\end{figure}

The spatially resolved Stokes vector of the generated beam is verified without having interacted with the sample, and measured again after interaction. 
Spatially dependent, single-shot polarization tomography using POVM measurements was initially demonstrated in \cite{Mustafa2022}, following the ideas outlined in \cite{Ling2006}. The experimental design for this involved the use of a Mach-Zehnder interferometer, however, the inherent sensitivity of a Mach-Zehnder interferometer to phase shifts was found to be a challenge experimentally, with frequent realignment required. 
Here, we present an updated experimental system for single-shot POVM measurements, incorporating a Sagnac interferometer, which provides significant benefits in terms of beam stability, allowing the possibility of polarization measurements over time, as shown in Figure~\ref{fig:SetUp}. 

The generated vector beam (Poincar\'e) first enters the system via a polarizing beam splitter (PBS), separating the horizontal and vertical polarization components such that they propagate in different directions around a split path Sagnac interferometer. Both the clockwise and counterclockwise interferometer path contains a glass slide for adjustment of the optical path length (see section~\ref{CalibrationSection}). Each path also contains a half waveplate ($\text{HWP}_1$ and $\text{HWP}_2$), which are rotated to control the ratio of horizontal to vertical polarization exiting each arm of the interferometer, with the rotation angles set to be $\theta_1 = \frac{1}{2}\sin^{-1}(a)$ and $\theta_2 = \frac{1}{2}\sin^{-1}(b)$ where $a$ and $b$ are the coefficients given in Eq.~\ref{Eq.ab}. This has the overall effect of transforming the initial PBS into a partially polarizing beam splitter (PPBS) with an output ratio of 21\%-79\% \cite{Ling2006}. 

One of the two output ports of the Sagnac interferometer is directed through an additional half waveplate ($\text{HWP}_3$) (with its fast axis set to $67.5^\circ$), the other passes through a quarter waveplate (with its fast axis set to $-45^\circ$) and a quartz plate for polarization dependent calibration (see section~\ref{CalibrationSection}). These two beam paths are then separated into their horizontal and vertical polarization components, using a second PBS, resulting in four beams corresponding to the POVM elements. 
The final beams are directed onto four quadrants of a single camera (see Figure~\ref{fig:SetUp}(a)), to allow a complete and spatially dependent polarization tomography in a single-shot measurement. 

\subsection{Experimental Calibration}
\label{CalibrationSection}

\begin{figure}
    \centering    \includegraphics[width=0.6\textwidth]{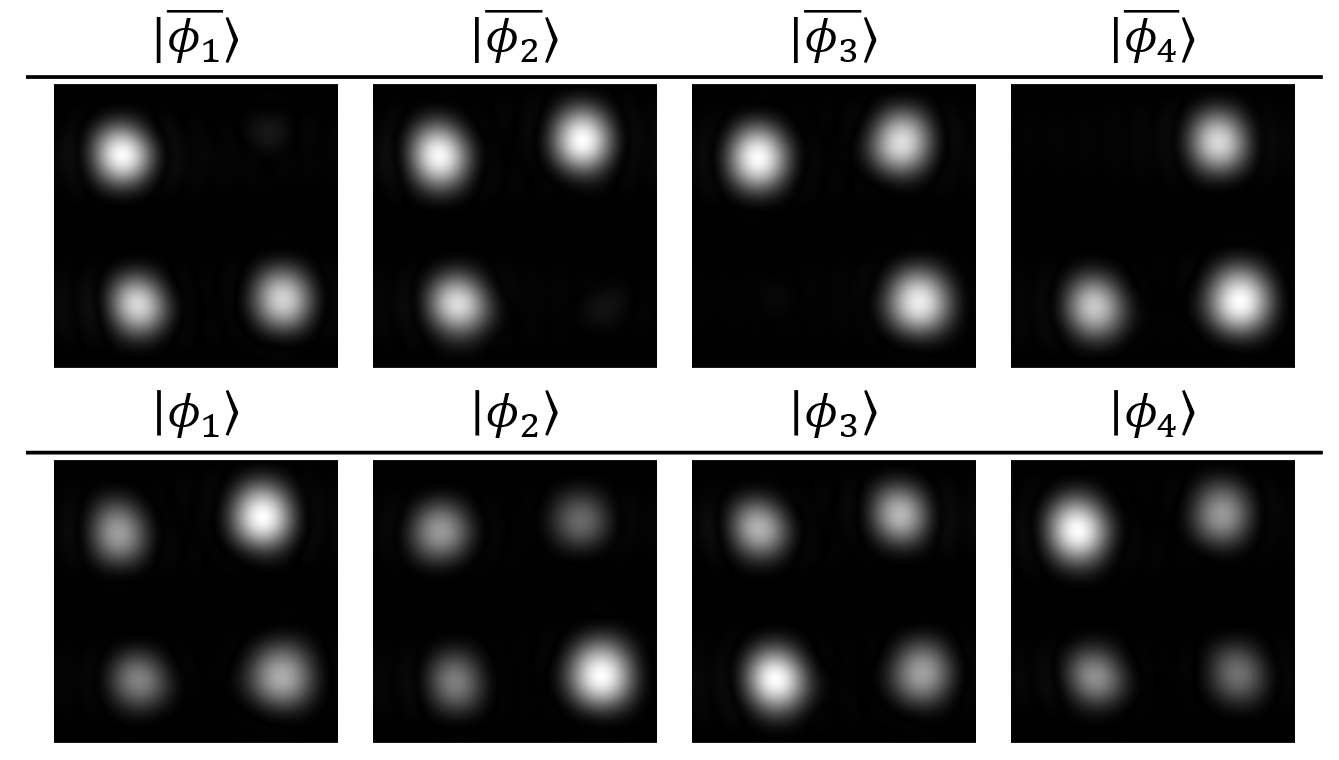}
    \caption{Experimental calibration measurements. Camera images of the homogeneously polarized orthogonal states, $\ket{\overline{\phi_i}}$ (top row) and POVM states, $\ket{\phi_i}$ (bottom row). The intensity measurements $I_i \ (i\in\{1,2,3,4\})$ are in the positions as indicated in the insert of Figure~\ref{fig:SetUp}.}
    \label{fig:Calibration}
\end{figure}

Using a POVM tomography based on a path displaced Sagnac interferometer eliminates optical path differences to first order and significantly improves resilience to common mode noise compared to previous designs \cite{Mustafa2022}. Nevertheless, due to the split path within the interferometer, there is a small difference in clockwise and anticlockwise optical path length, which requires periodic calibration and measurements of the instrumentation matrix $\Pi$, which we detail in the following.

Homogeneously polarized beams with polarizations orthogonal to the POVM states are particularly sensitive to any misalignment and hence are used to calibrate the experiment and also to test stability (see section \ref{sec stability} ). These orthogonal states, $\ket{\overline{\phi_i}}$, with $\langle\overline{\phi_i}|\phi_i \rangle=0$ are given by
\begin{equation}
  \begin{split}
    \ket{\overline{\phi_1}} = b\ket{h}-a\ket{v}, \quad \ \ket{\overline{\phi_2}} = b\ket{h}+a\ket{v}, \\
    \ket{\overline{\phi_3}} = a\ket{h}-ib\ket{v}, \quad \ \ket{\overline{\phi_4}} = a\ket{h}+ib\ket{v}.
  \end{split}
  \label{Eq.MinusStates} 
\end{equation}
As $P_i=\bra{\overline{\phi_i}} \hat{\pi}_i \ket{\overline{\phi_i}}=0$, measurement of each state $\ket{\overline{\phi_i}}$ should result  in $I_i = 0$, and equal intensities in the other 3 camera quadrants for an ideal interferometer. Monitoring and minimising the light level observed for the orthogonal states therefore allows us to calibrate the instrument. 
Calibration aims to eliminate any phase difference between the clockwise and anti-clockwise arm of the interferometer, for both horizontal and vertical polarization components, which is achieved by tilting one of  the glass slides within the interferometer and the quartz plate (QP) (see Figure~\ref{fig:SetUp}) . 
In theory, a single glass slide in one of the interferometer arms is sufficient to compensate for path differences, however, we opted for a symmetric setup to balance any intensity loss. 
To calibrate the system we start by tilting only one of the glass slides until the $I_3$ ($I_4$) camera quadrant records minimum intensity when measuring $\ket{\overline{\phi_3}}$ $\left(\ket{\overline{\phi_4}}\right)$, while the other glass slide is kept stationary, at normal incidence. Afterwards, the quartz plate is tilted to ensure minimum intensity in the $I_1$ ($I_2$) camera quadrant for $\ket{\overline{\phi_1}}$ $\left(\ket{\overline{\phi_2}}\right)$. 
Experimental images of the orthogonal states after calibration can be seen in the top row of Figure~\ref{fig:Calibration}. 

Once the system is calibrated, an experimental instrumentation matrix $\Pi$ can be obtained. From Eq.~\ref{Eq.Projection}, we know that $\Pi$ relates normalized Stokes vectors to the normalized output intensities. 
We generate the 4 POVM states (Eq.~\ref{Eq.POVMStates}) as the input beams, and record the resulting intensity profiles, as shown in the bottom row of Figure~\ref{fig:Calibration}. The normalized intensity four-vector $\vec{I}_i^{(\text{N})}$ is found by recording the intensity within each quadrant and dividing it by the total intensity incident on the camera.  
From the experimentally recorded $\vec{I}_i^{(\text{N})}$ and the ideal Stokes vectors $\vec{S}^{\rm (N)}$, the experimental $\Pi$ can simply be found by inverting Eq.~\ref{Eq.Projection}. The calibration method is sensitive to any discrepancies in beam generation, and care has to be taken to ensure high fidelity operation of the DMD-based vector beam generation. 
An example of an experimentally measured instrumentation matrix is,

\begin{equation}
    \Pi_{\text{exp}} = \frac{1}{4}
    \begin{pmatrix}
      0.950 & 0.826\sqrt{1/3} & 0.786\sqrt{2/3} & -0.015\\
      1.041 & 0.867\sqrt{1/3} & -0.804\sqrt{2/3} & -0.017\\
      0.955 & -0.810\sqrt{1/3} & 0.007 & -0.875\sqrt{2/3}\\
      1.053 & -0.883\sqrt{1/3} & 0.008 & 0.915\sqrt{2/3}
    \end{pmatrix}.
    \label{Eq.ExpInstrument2}
\end{equation}

Comparison with the ideal instrumentation matrix Eq.~\ref{Eq ideal Pi} shows some discrepancies, indicating imperfections of the optical components and the alignment. While the experimental setup transforms the input state onto slightly different polarization states than the intended tetrahedral states, polarization tomography can still be performed as long as a mapping between the Stokes vectors and measured POVM states exists and there is a large enough coverage of the Poincar\'e sphere. 

\subsection{Investigation of experimental stability \label{sec stability}}

As described in section \ref{sec setup}, we have developed a Sagnac-based interferometer to perform one-shot generalized measurements for polarization tomography. Performing Mueller matrix measurements, especially when determining the time evolution of optical activity e.g. of bio-chemical samples, requires the system to be optically stable during the entire evaluation process, without the need of re-calibration.

\begin{figure}
    \centering
    \includegraphics[width=0.7\textwidth]{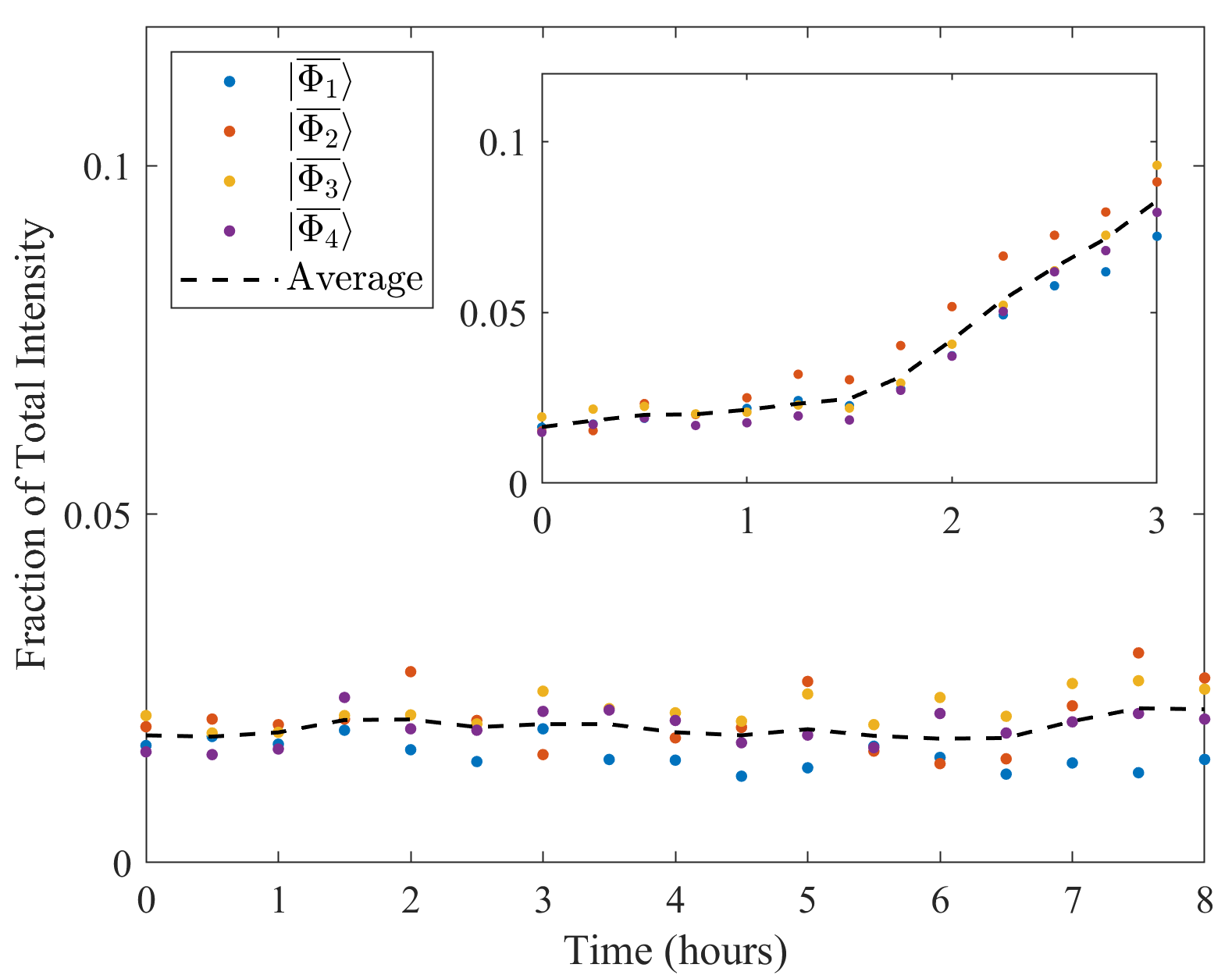}
    \caption{Testing stability of the Sagnac interferometer in terms of the alignment states. The data points show  $I^{\rm (N)}$ in camera quadrant $I_i$ for the measured orthogonal state $\ket{\overline{\phi_i}}$, where $i\in\{1,2,3,4\}$, and the black dashed line shows a smoothing spline fit to their average. The main plot shows data taken at 30 minute intervals over a total time of 8 hours during a quiet time in the laboratory. Stability decreases when there is a lot of traffic in the laboratory, as indicated in the inset, showing data taken at 15 minute intervals over 3 hours. }
    \label{fig:Stability}
\end{figure}

In this section, we briefly outline an investigation performed to asses the stability of the experimental setup. Just like for the experimental calibration we make use of the orthogonal states $\ket{\overline{\Phi_i}}$, which in theory should produce zero intensity in the corresponding camera quadrant as discussed in section \ref{CalibrationSection}.
If the alignment of the interferometer drifts over time, intensities in these quadrants are expected to increase. Our experimental findings are summarized in Figure~\ref{fig:Stability}, showing $P_i=\bra{\overline{\phi_i}} \hat{\pi}_i \ket{\overline{\phi_i}}$, measured as the fraction of the total intensity in the corresponding quadrant. The main figure shows measurements taken over 8 hours at 30 minute intervals in a quiet lab, and in the inset over 3 hours at 15 minute intervals during times of intense activity and traffic next to the experiment. After initial alignment and calibration, we typically achieve probabilities as low as $P_i \approx 2 \% $. In a quiet laboratory, calibration quality changes only insignificantly over the whole 8 hour period, whereas in a busy laboratory calibration deteriorated markedly. We attribute this gradual misalignment to slight shifts of the glass slides due to vibrations in the laboratory. This may be further suppressed in a custom-built monolithic interferometer setup, however, for the measurements reported in this paper the stability is entirely sufficient.

\section{Mueller matrix results \label{sec Mueller Results}}

As outlined in section \ref{sec Mueller Theory}, the Mueller matrix of a sample with a uniform optical activity can be obtained by measuring its effect on a Poincar\'e beam. Experimentally, this requires us to identify the Stokes parameters of the beam itself (without the sample) and after interaction with the sample. 

Here, we choose a beam generated from a superposition of horizontally and vertically polarized Laguerre Gaussian ($\text{LG}_p^\ell$) modes, of the form, 
\begin{equation}
    \ket{\Psi} = \text{LG}_1^0\ket{h}+\text{LG}_0^2\ket{v},
    \label{Eq.PoincareBeam}
\end{equation}
where $p$ and $\ell$ are the radial and azimuthal mode numbers, respectively. As the mode numbers of both polarization contributions are identical, $N=2p+|\ell|=2$, they have the same Gouy phase so that the beam maintains its shape upon propagation apart from an overall change of size. The experimentally measured polarization profile $\cal{S}_{\rm in}$ of $\ket{\Psi}$ is shown in the top left of Figure~\ref{fig:MuellerBeams}, with the simulated ideal polarization profile displayed below.

\begin{figure}
    \centering    \includegraphics[width=0.95\textwidth]{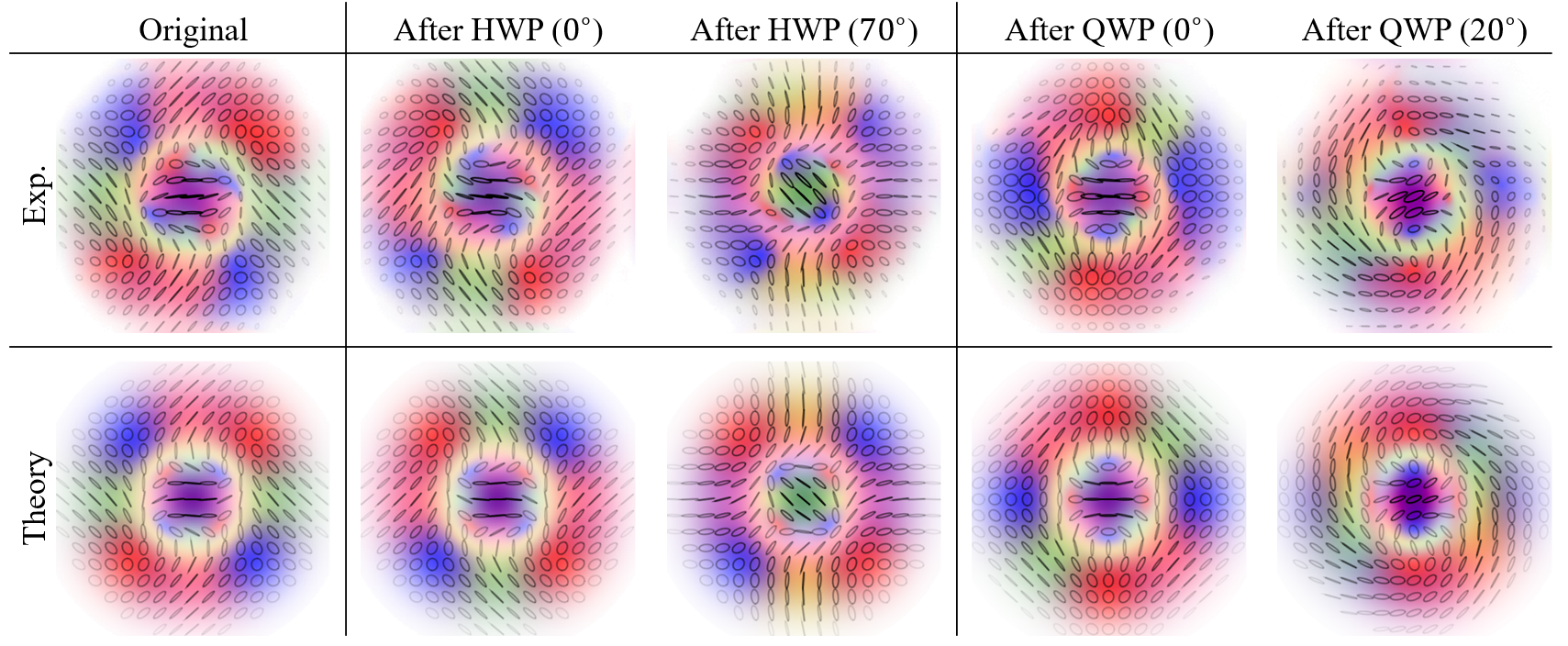}
    \caption{Experimentally measured with the Sagnac interferometer (top) and theoretical (bottom) polarization profiles $\cal{S}_{\rm in}$ of a $\text{LG}_1^0\ket{h} + \text{LG}_0^2\ket{v}$ Poincar\'e beam before (left) and $\cal{S}_{\rm out}$ after passing through either a half waveplate or quarter waveplate with their fast axis rotated with respect to the horizontal to the angle indicated in the figure. The color scheme is described in Figure~\ref{fig:POVMStates}.
    }
    \label{fig:MuellerBeams}
\end{figure}

The sample to be analyzed is placed in the beam path, before the beam enters the interferometric system, as indicated by the dashed box in Figure~\ref{fig:SetUp}, and the resulting Stokes parameter $\cal{S}_{\rm out}$ is obtained as before. The Mueller matrix of the sample can then be obtained from Eq.~\ref{Eq.SolveMueller}.
The full profile of the Poincar\'e beam intensities and hence Stokes vectors provide an overcomplete measurement, and we can therefore afford to eliminate camera pixels at low intensities that may be compromised by noise -- as long as the remaining sections of the Poincar\'e beam still span the full polarization state space. Specifically, we choose to include only those camera pixels in the analysis with an intensity greater than 5\% of the peak intensity.

\subsection{Mueller matrix measurement of rotated waveplates}

We illustrate and test the proposed method by measuring the Mueller matrices of half and quarter waveplates at various angles, for which the theoretical matrices are readily available as, 
\begin{equation}
    M_T(\delta,\theta) = \begin{bmatrix}
        1 & 0 & 0 & 0\\
        0 & \cos^2(2\theta)+\sin^2(2\theta)\cos(\delta) & \cos(2\theta)\sin(2\theta)(1-\cos(\delta)) & \sin(2\theta)\sin(\delta)\\
        0 & \cos(2\theta)\sin(2\theta)(1-\cos(\delta)) & \cos^2(2\theta)\cos(\delta) + \sin^2(2\theta) & -\cos(2\theta)\sin(\delta) \\
        0 & -\sin(2\theta)\sin(\delta) & \cos(2\theta)\sin(\delta) & \cos(\delta)
    \end{bmatrix},
    \label{Eq.TheoryMatrix}
\end{equation}
where $\delta$ is the phase difference between the fast and slow axis and $\theta$ is the angle of the fast axis with respect to the horizontal. For a HWP and QWP, $\delta=\pi$ and $\pi/2$, respectively. A selection of the measured polarization profiles of the chosen Poincar\'e beam after passing through the waveplates at different angles of the fast axis are shown in Figure~\ref{fig:MuellerBeams} along with the corresponding theoretical plots.

\begin{figure}
    \centering    \includegraphics[width=1\textwidth]{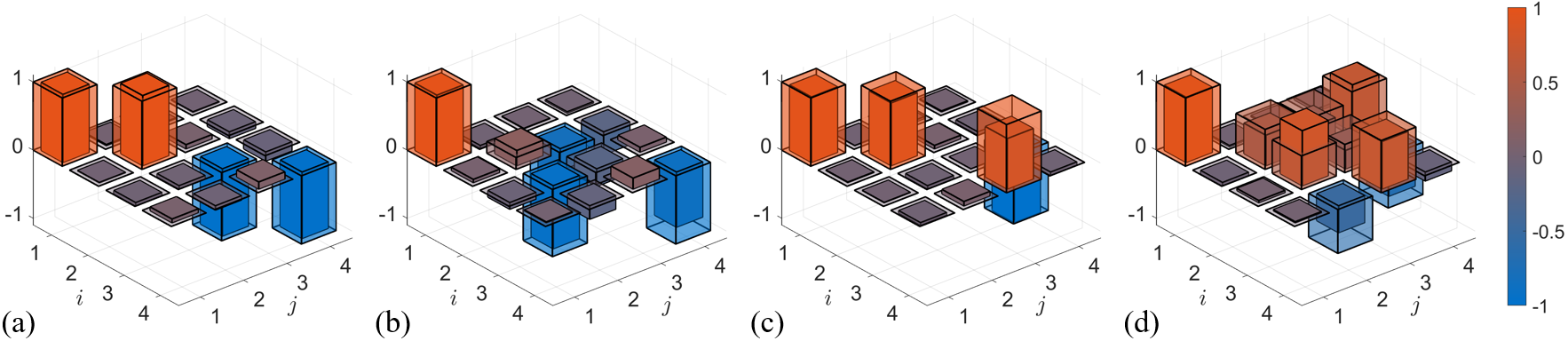}
    \caption{Comparison of experimental and theoretical Mueller matrices. The bar graphs show the $m_{i,j}$ entries of Mueller matrices where $i,j \in \{1,2,3,4\}$ are the row and column indices, respectively. Theoretical values are indicated by the transparent columns and experimental values given by narrower opaque columns, with positive and negative values are indicated by a red-to-blue color gradient. Examples are given for a HWP with fast axis at $0^\circ$ (a) and $70^\circ$ (b) and a QWP at $0^\circ$ (c) and $20^\circ$ (d).  
    }
    \label{fig:BarCharts}
\end{figure}

From the measured polarization profiles, the Mueller matrices could be obtained as described above. Figure~\ref{fig:BarCharts} compares a representative selection of the measured matrices to the theoretically expected matrices for the same fast axis angles as shown in Figure~\ref{fig:MuellerBeams}. The bar charts in Figure~\ref{fig:BarCharts}(a) and (b) show results for the HWP and \ref{fig:BarCharts}(c) and (d) show the results for the QWP, where theoretical values are given as transparent columns, measured values as narrower opaque columns.

\begin{figure}
    \centering 
    \begin{subfigure}[b]{0.49\textwidth}
        \centering
        \includegraphics[width=0.99\textwidth]{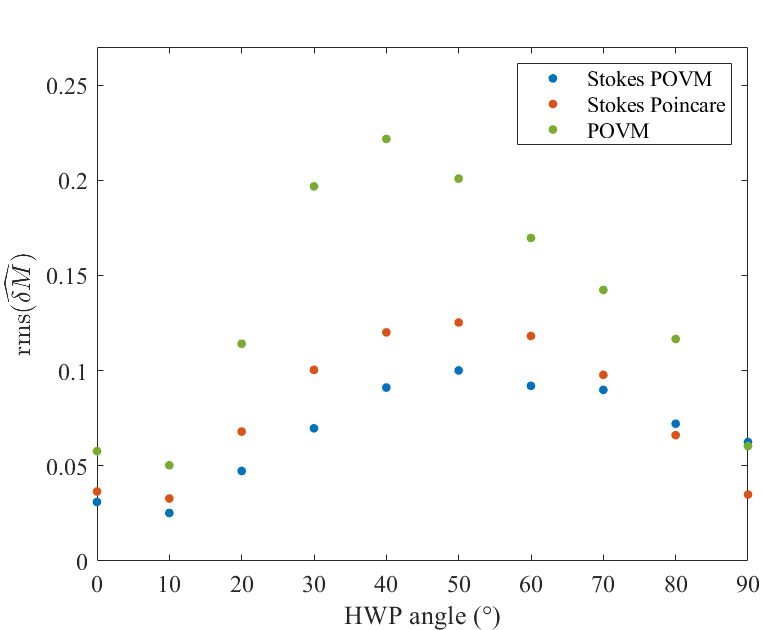}
        \caption{}
    \end{subfigure}%
    ~
    \begin{subfigure}[b]{0.49\textwidth}
        \centering
        \includegraphics[width=0.99\textwidth]{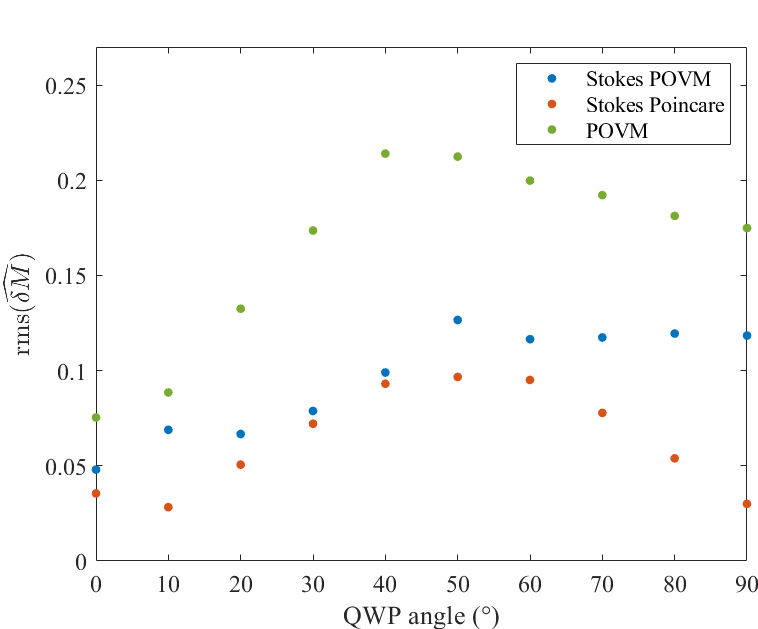}
        \caption{}
    \end{subfigure}
    \caption{Comparison of calculated errors of the measured Mueller matrices of a HWP (a) and QWP (b) for the different methods used. Green data points show errors for the single shot POVM measurements, blue and orange data points are the errors when Stokes measurements are used either using the 4 POVM states as input beams (blue) or a Poincar\'e beam (orange). The errors are calculated using Eq.~\ref{Eq.rmserror}}
    \label{fig:ErrorComparison}
\end{figure}

An error value, quantifying the discrepancy between the Mueller matrices obtained by our single-shot POVM measurements and the theoretically predicted quantities is shown as green points in Figure~\ref{fig:ErrorComparison}. The error is evaluated 
as root mean square (rms) average \cite{Suarez-Bermejo2022},
\begin{equation}
    \text{rms}(\widehat{\delta M}) = \frac{1}{4}\sqrt{\sum_{i,j=1}^4|\delta M_{ij}|^2},
    \label{Eq.rmserror}
\end{equation}
where, $\widehat{\delta M} = \widehat{\delta M_E} - \widehat{\delta M_T}$ is the difference between the experimentally measured Mueller matrix and the theoretically expected matrix, and $i$ and $j$ are the row and column indices. We note that the obtained error also includes small uncertainties in determining the exact rotation angle of the retardation plates $\theta$ and rely on our knowledge of the parameter $\delta$ as given by optics manufacturers.

We also asses the performance of our single-shot POVM device compared to sequential Stokes analysis by analyzing the light in the $\ket{h},\ \ket{v},\ \ket{d},\ \ket{a},\ \ket{r}$ and $\ket{l}$ polarization basis using a rotating wave plate setup (consisting of a rotating QWP, HWP and stationary polarizer) for the sample as above.    
The sequential Mueller matrix analysis was performed in two ways, performing Stokes analysis of a Poincar\'e probe beam, and of four homogeneous POVM beams, shown as orange and blue data points in Figure~\ref{fig:ErrorComparison} respectively. In each case, measurements are taking before and after the optical sample.

For the Stokes Poincar\'e Mueller analysis, we used the Poincar\'e beam of Eq.~\ref{Eq.PoincareBeam} as probe, requiring a total of 6 sequential measurements per sample. This results in spatially resolved Stokes parameters associated with each camera pixel, from which the Mueller matrices could be calculated using Eq.~\ref{Eq.SolveMueller}, analogous to data processing for the single-shot measurements. 

For the Stokes POVM Mueller analysis, the probe beam was cycled through four uniformly polarized beams corresponding to each of the POVM polarization states in Eq.~\ref{Eq.POVMStates}, requiring a series of 24 ($4\times6$) consecutive images per sample. Stokes parameters were acquired for each beam by averaging over the entire intensity profiles, resulting in four input and output Stokes vectors, allowing us to construct two $4\times4$ matrices, and solve $\vec{S}_{\rm out} = \vec{M}\,\vec{S}_{\rm in}$ directly.

Our comparison in Figure \ref{fig:ErrorComparison} shows that the Mueller matrices obtained with our single-shot POVM method have an error of similar magnitude (within a factor of two) as those measured with traditional Stokes polarimetry, despite the large gain in measurement speed. It is not surprising that Stokes based analysis performs slightly better, given that it consists of a simple sequence of three optical elements compared to the Sagnac setup shown in Figure \ref{fig:SetUp}. The similar wavelength dependence of all three measurement techniques, with a maximal error at $40^{\circ}$, points to a slight alignment error, potentially causing a slight tilt of the waveplates relative to the beam path.

All of the previous results were derived from stationary images, however, as all the information required for our Mueller matrix measurements can be acquired in a single-shot, it is possible to take dynamic measurements, measuring changes to optical activity in time. To give an example of this, we placed a HWP in the sample location that was mounted in a rotation mount controlled by a stepper motor. A live video could then be recorded of the beam passing through the HWP while it rotated a full $360^\circ$. 
The raw video can be seen in Visualization 1 along with the reconstructed polarization profile.
As the stepper motor rotated a full $360^\circ$ in 1.13 seconds and the camera recorded at 39.68 frames per second, each frame of the video shows the altered intensity distributions after the HWP had advanced by $8^\circ$. Here, we are clearly only limited by the frame rate of the camera, and employing a faster camera would provide more time resolution. In Visualization 1 we show the recorded video slowed down by a factor of 10, in order to clearly view the changes to the recorded intensity and reconstructed polarization profiles.

\subsection{Measuring the retardance of a tilted quarter waveplate}
\label{TiltedQWPSection}

As an example of how we can use these Mueller matrix measurements, we measure the retardance of a tilted QWP, with its fast axis set to $0^\circ$ and its normal vector tilted by an angle $\alpha$ with respect to the beam propagation as indicated in Figure~\ref{fig:SetUp}(b). 

Birefringent materials impart a phase shift between two orthogonal polarization components of incident polarized light. As such, they are commonly used for polarization modulation, most frequently in the form of QWPs and HWPs. 
This phase shift (or retardance) depends on the properties of the material, such as the thickness, and the ordinary ($n_o$) and extraordinary ($n_e$) refractive indices at the relevant wavelength. 
Birefringent waveplates are designed for use at normal incidence, however, the retardance changes depending on the angle of incidence (AOI) of the incident light, due to the increased distance traveled through the material. The optical path difference between the ordinary and extraordinary rays changes, imparting an unintended phase shift between the rays. 
The significance of this change in retardance with AOI is material dependent: Polymer waveplates, for example, are fabricated to allow stability over a range of AOI, whereas waveplates made of crystalline quartz incur greater retardance changes, even for small AOI \cite{ThorlabsQWP}.

The Mueller matrix of a waveplate with its fast axis at $0^\circ$ and arbitrary retardance ($\delta$) is, 
\begin{equation}
    M_T(\delta, \theta=0) = \begin{bmatrix}
        1 & 0 & 0 & 0 \\
        0 & 1 & 0 & 0 \\
        0 & 0 & \cos(\delta) & -\sin(\delta) \\
        0 & 0 & \sin(\delta) & \cos(\delta)
    \end{bmatrix}.
    \label{Eq.TheoryMatrix2}
\end{equation}

For a matrix of the form of Eq.~\ref{Eq.TheoryMatrix2}, the correct sign and magnitude of the retardance can be obtained, simply using the fact that $\delta = \tan^{-1}(\sin(\delta)/\cos(\delta))$.  
Therefore, from a measured Mueller matrix, there are 4 ways to compute the retardance,
\begin{equation}
    \delta_1 = \tan^{-1}\left(-\frac{m_{34}}{m_{33}}\right), \ \delta_2 = \tan^{-1}\left(-\frac{m_{34}}{m_{44}}\right), \
    \delta_3 = \tan^{-1}\left(\frac{m_{43}}{m_{33}}\right), \ \delta_4 = \tan^{-1}\left(\frac{m_{43}}{m_{44}}\right),
    \label{Eq.RetardanceEquations}
\end{equation}
where, $m_{i,j}$ are the matrix elements of $M_T$ with row and column indices $i$ and $j$ respectively.

To test our Mueller matrix polarimetry procedure for measuring retardance, we used a multi-order, crystalline quartz quarter waveplate (ThorLabs WPMQ05M-633), mounted on a rotation stage to allow varying incident angles. With its fast axis set to $0^\circ$, the same Poincar\'e beam used previously is used at the test beam, and the Mueller matrix is found using the same procedure as before, for incident angles ranging from $0^\circ$ to $40^\circ$.

\begin{figure}
    \centering
    \includegraphics[width=0.7\textwidth]{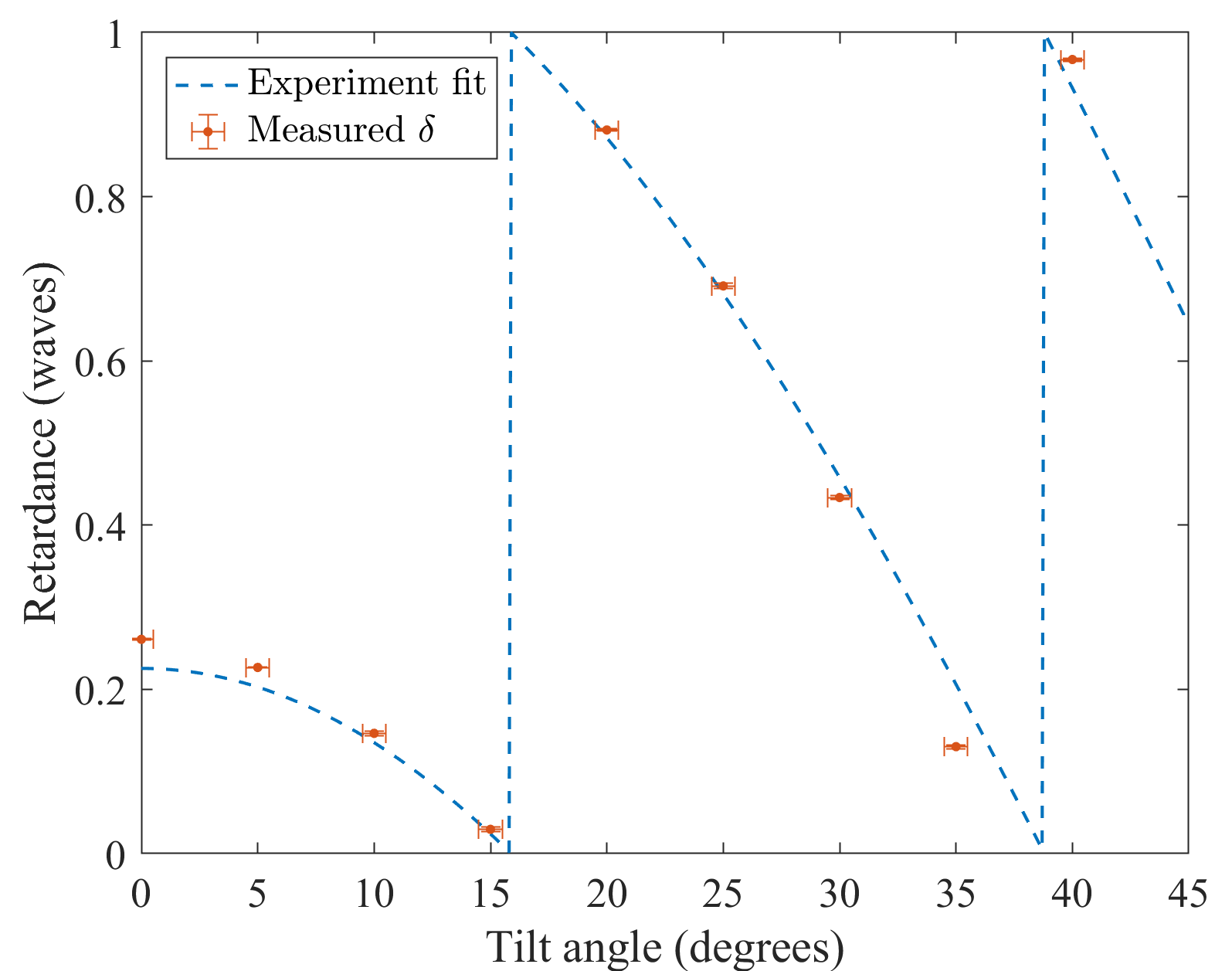}
    \caption{Retardance against angle of incidence for a QWP with its fast axis at $0^\circ$. Measured retardance from Mueller matrix measurements are shown as orange data points with the corresponding fit given as a dashed blue line. }
    \label{fig:TiltedQWP}
\end{figure}

The retardance of the tilted QWP could then be obtained for each incident angle, using the last 4 entries of the measured Muller matrix and Eq.~\ref{Eq.RetardanceEquations}, leading to 4 separate values.
These values should all be the same, but due to experimental errors, there are slight deviations between them. To get a final value for the retardance, we find the average of $\delta_1$, $\delta_2$, $\delta_3$ and $\delta_4$ and compute the standard deviation for an associated error. 
The results of the measured $\delta$ are shown in Figure~\ref{fig:TiltedQWP} as orange data points.

To determine a fit to the measured retardance, we follow a similar procedure as outlined in \cite{Gu2018}, but modified for a multi-order waveplate rather than a zero-order waveplate, as used in the mentioned paper. The theoretical retardance can therefore be calculated as,
\begin{equation}
    \delta_T(\theta,\alpha) = \frac{2\pi T}{\lambda}\left(\sqrt{n_e^2 - \frac{n_e^2\cos^2(\theta) + n_o^2\sin^2(\theta)}{n_o^2}\sin^2(\alpha)} - 
    \sqrt{n_o^2 -\sin^2(\alpha)} \right),
    \label{Eq.TheoryDelta}
\end{equation}
where $\theta$ is the angle of the waveplates fast axis, $\alpha$ is the angle of incidence, $T$ is the thickness of the waveplate, $n_o$ and $n_e$ are the ordinary and extraordinary refractive indices and $\lambda$ is the wavelength. In our case, $\lambda=632.8\text{ nm}$, and the relevant refractive indices of crystalline quartz at this wavelength are $n_o=1.543$ and $n_e=1.552$ \cite{Crystran} and the thickness can be found from the manufacturers website \cite{ThorlabsQWP}.  

For a waveplate with its fast axis at $0^\circ$, Eq.~\ref{Eq.TheoryDelta} reduces to, 
\begin{equation}
    \delta_T(\theta=0,\alpha) = \frac{2\pi T}{\lambda}\left(\sqrt{n_e^2\left(1 - \frac{\sin^2(\alpha)}{n_o^2}\right)} - 
    \sqrt{n_o^2 -\sin^2(\alpha)} \right).
    \label{Eq.TheoryDelta2}
\end{equation}
To obtain $\delta_T$ in units of waves, as shown by the blue line in Figure~\ref{fig:TiltedQWP}, we simply apply the modulo operation, to get the result in the range $[0,2\pi]$ and divide by $2\pi$. We note however that the values for $n_e$ and $n_o$ available from the manufacturer are only known to 4 significant figures.

We have therefore carried out a fit to the experimentally measured retardance using Eqs.~\ref{Eq.TheoryDelta2}, determining the refractive indices as $n_0 = 1.54302 \pm 0.00032$ and $n_e = 1.55199 \pm 0.00033$
(which are similar to the expected values quoted above). We additionally determined the thickness of the material to be $T= 1.0023\pm0.0001$ mm. The fit is shown in Figure~\ref{fig:TiltedQWP} as a dashed orange line. 

\section{Conclusion}

Mueller calculus is a powerful technique for describing the action of optical elements or optically active media on the polarization of a light beam. Identifying the Mueller matrix of a sample determines its retardance magnitude and orientation, and allows us to quantify linear and circular birefringence and dichroism, which in turn characterize magneto-optical effects as well as optical activity in chiral media. Here we have demonstrated a single-shot method to determine Mueller matrices, based on the minimum number of (generalized) measurements. These POVM measurements are realized in a displaced Sagnac interferometer, achieving an interferometric stability of better than 2$\%$ in a one hour time period. Spatially resolved measurements in conjunction with a Poincar\'e probe beam then allow the time-resolved identification of Mueller matrices. We have tested our device for stationary rotated and tilted retardation plates, obtaining excellent agreement with theoretical predictions at comparable errors to conventional rotating wave plate measurements. We have furthermore demonstrated the viability of measuring dynamic processes by recording video data during controlled rotation and tilting of the retardation plates. While our suggested method is based on generalized measurements of symmetric states, and hence treats all polarization states equally, the sensitivity to detect specific polarization responses could be enhanced by choosing suitable non-symmetric tetrad states.

We expect our method to provide a convenient alternative approach for studying real-time monitoring of rapid optical activity changes due to biological phenomena, physical or chemical reaction processes, and complex fluid studies, but equally for the long-time non-invasive investigation of slow biochemical processes relevant e.g. for monitoring in agriculture and food industries. Our method is wavelength independent, and its temporal resolution is only limited by the video refresh rate of the camera, promising resolutions in the range of nanoseconds with commercial scientific cameras.

\begin{backmatter}
\bmsection{Funding}

A.M. acknowledges financial support from the UK Research and Innovation Council via grant EPSRC/DTP
2020/21/EP/T517896/1.
M.A.A. received funding via QuantIC (EP/M01326X/1) and Fraunhofer CAP, S.P. acknowledges funding through SUPA (Scottish Universities Physical Alliance) Distinguished Visitors Programme, Conselho Nacional de Desenvolvimento Científico e Tecnológico, Instituto Nacional de Ciência e Tecnologia de Informação Quântica and Coordenação de Aperfeiçoamento de Pessoal de Nível Superior (CAPES-PRINT).

\bmsection{Acknowledgments}
We are grateful to Dr Sarah Croke for many insightful discussions in the crucial early stages of devising this work.

\bmsection{Disclosures}
The authors declare no conflicts of interest.

\bmsection{Data availability}
Data underlying the results presented in this paper are not publicly available at this time but may be obtained from the authors upon reasonable request.

\bmsection{Supplemental document}
See Visualization 1 for supporting content. 

\end{backmatter}



\end{document}